# The Effect of Alternating Magnetic Field on Acceleration of Charged Particles

Olga Medvedeva

Institute for Spectroscopy RAS, Moscow, Troitsk, Russia

Medolg83@gmail.com;

The self-consistent problem of the wave and particle spectrum is formulated and solved for acceleration of particles in a homogeneous magnetic field that varies periodically with time. It follows from the obtained solutions that when account is taken of the synchrotron radiation, the diffusion coefficient Do of ultrarelativistic electrons does not differ from the Fermi value. An expression is obtained for the minimum concentration of the accelerated particles, at which the cyclotron instability ensures the scattering necessary for effective acceleration.

## 1. Introduction

If a slow periodic time variation of a uniform magnetic field B in a turbulent plasma is accompanied by conservation of the adiabatic invariants $p_z = const$, $p_\perp{}^2 / B = const$ ( $p_\perp$ and $p_z$ are the particle-momentum components perpendicular and parallel to the field), then the total momentum p of the charged particles can increase exponentially relative to the time t as a result of betatron acceleration and non-adiabatic scattering by the hydromagnetic turbulence (the Alfven magnetic pumping) [l-4]. In a denser plasma, the role of the scattering can be assumed by particle collisions [4]. In [4] the case of harmonic variation of the field

$$B = B_0(1 + \beta \cos Wt), \quad \beta < 1 \qquad (1)$$

has considered

In [5] was shown that the kinetic equation for the distribution function $\bar{f}(p,t)$ averaged over the fast turbulent pulsations can be reduced in the quasilinear approximation, in the presence of a force $F = 1/2(p_\perp / B)dB / dt$, to the diffusion equation in momentum space, with a diffusion coefficient D( p) that depends on the parameters of the alternating magnetic field and on the turbulence spectrum $\Phi(k) = dh^2 / dk$ ($h^2$ is the turbulence intensity and k is the wave number. As is well known [6,7], in the diffusion approximation the turbulence spectrum determines the rate of growth (or damping) of the plasma oscillations, which in turn depends on

$\bar{f}(p,t)$, so that the problem of finding the spectrum of the waves and of the particles should be formulated in a self-consistent manner.

In the present paper we obtain, by successive approximations, a solution of the self-consistent problem without using numerical methods, for nonrelativistic and ultrarelativistic stationary and nonstationary accelerations, and also for ultrarelativistic stationary acceleration of electrons with allowance for the synchrotron radiation. It is shown that at the plasma instability boundary the turbulence spectrum has a universal form $\Phi(k) = \Phi_0 / k^2$. We compare the obtained diffusion coefficient with the Fermi coefficient, and also with the diffusion coefficient for turbulent acceleration, as calculated in [8,9]. We show that the Alfven acceleration leads in the presence of synchrotron radiation to the same particle spectrum, which is formed according to [8] under Fermi acceleration. In Sec. 3 we investigate the limit of applicability of the statistical Alfven acceleration with allowance for the synchrotron radiation.

All the results are obtained for a collisionless plasma under the following assumptions: 1) the variation of the magnetic field is given by Eq. (1); 2) the plasma is assumed to be cold and the fraction of accelerated particles small; 3) the velocities of the fast particles exceed the Alfven velocity cA; 4) the main scale of turbulence L is much larger than the Larmor radii of the fast particles (in this case L does not exceed $L \sim Tc_{A}$, where $T = 2\pi / W$; 5) the angular distribution of the wave vectors of the pulsations is isotropic or has a maximum at small angles to the field, so that the main contribution to the scattering is made by waves traveling along B.

## 2. Solution of Self-Consistent Problem

We shall solve Eqs. (14), (15), and (19) of [5] by successive approximations, assuming as the zeroth approximation a wave spectrum in the form $\Phi(k) = \Phi_0 / k^2$, at which the undetermined angle interval in (15) of [5] is given by

$$\int \frac{\cos\theta \sin\theta d\theta}{|P(\theta)|} = \frac{\omega_B}{\Phi_0 cp} \left( C_0 - b|\cos\theta| \right),$$
$$b = \begin{cases} +1, & 0 < \theta < \pi / 2 \\ -1, & \pi / 2 < \theta < \pi. \end{cases} \tag{2}$$

We obtain the integration constant $C_0$ from Eq. (6) of [5], which is transformed with the aid of (10), (11) of [5], and (2) into

$$\int_0^\pi \sin\theta \left( C_0 - b|\cos\theta| \right) d\theta = 0,$$

from which it follows that $C_0 = 1/2$. Substituting (20) (with $C_0 = 1/2$) in (15) of [5] and integrating with respect to the angle, we obtain

$$D(p) = D_0 \varepsilon p, \quad D_0 = \frac{m^2 c^2 \omega_B^2 p^2}{36 e^2 c \Phi_0} \left\langle \left( \frac{\dot{B}}{2B} \right)^2 \right\rangle, \qquad (3)$$

so that Eq. (14) of [5] takes the form

$$\frac{\partial \overline{f}}{\partial t} = \frac{D_0}{p^2} \frac{\partial}{\partial p} \left[ p^2 \varepsilon \frac{\partial \overline{f}}{\partial p} \right]. \qquad (4)$$

To find the next approximation for $\Phi(k)$, we must substitute the solution of (4) in (19) of [5]. In the nonrelativistic case ($\varepsilon = 1$) the solution of (4) is

$$\overline{f} = \frac{\partial \overline{f} A}{(D_0 t)^2} \exp\left( -\frac{p}{D_0 t} \right),$$

and in the ultrarelativistic case ($\varepsilon = p$), the nonstationary solution of (4), as shown in [12], is

$$\overline{f} = \frac{A}{p^{1/2}} \exp\left( -\frac{\ln^2 p}{4 D_0 t} \right).$$

In the particular case of stationary acceleration ($\partial \overline{f} / \partial t = 0$), Eq. (4) with $\varepsilon = 1$ is satisfied with the function $\overline{f} \sim p^{-2}$, and at $\varepsilon = p$ by the function $\overline{f} \sim p^{-3}$. It is easily shown that in each of these four particular cases we have

$$\lim_{p \to \infty} \left( \frac{\omega_B^2}{c^2 k^2} - p^2 \right) \overline{f}(p) / 2 \int_{\omega_B/ck}^\infty p \overline{f} dp = 0,$$

so that expression (19) of [5] takes the universal form

$$\Phi(k) = \mp \frac{m^2 c^2 \omega_B}{3 e^2 \omega k} \frac{\dot{B}}{2B} \qquad (5)$$

and coincides with the initial approximation $\Phi(k) = \Phi_0 / k^2$, for Alfven waves whose dispersion equation is $\omega = cAk$. Thus, the indicated solutions of (4) together with (3) and (5) (Alfven waves) are exact solutions of the self-consistent problem for the cases considered above.

According to (5), (3) of [5] and (1), the final expressions for the Alfven-wave spectrum and the diffusion coefficient $D_0$, recognizing that $\dot{B} / 2B \approx -1/2\beta W \sin Wt$ and putting $|\sin Wt| \approx \left\langle \sin^2 Wt \right\rangle^{1/2} = 1/\sqrt{2}$, take the form

$$\Phi(k) = \frac{\Phi_0}{k^2}, \quad \Phi_0 = \frac{B_0{}^2 \beta W}{6\sqrt{2} c_A}, \tag{6}$$

$$D_0 = \frac{\sqrt{2}}{48} \beta W \frac{c_A}{c}. \tag{7}$$

It follows from (6) that the energy of the turbulent pulsations

$$\frac{h^2}{8\pi} = \frac{1}{8\pi} \int \Phi(k)dk = \frac{\beta W B_0{}^2}{48\sqrt{2\pi\omega}}$$

is much smaller than the magnetic energy $B_0{}^2 / 8\pi$, since $\beta < 1$, and the resonant frequencies are $\omega >> W$. Thus, the condition $h^2 << B_0{}^2$ for quasilinearity [7] of the initial kinetic equation (2) of [5] at the plasma stability limit is well satisfied.

In [8], the following expression is obtained for the Fermi diffusion coefficient $D_F$ in momentum space, as applied to the case of the most effective Fermi acceleration (reflection from long strong waves, when $\Phi(k) \sim k^{-v}$, v>2):

$$D_F \approx \frac{c_A{}^2}{Lc} \frac{h_m}{B_0}, \tag{8}$$

where L is the main scale, and $h_m = B_m - B_0$ is the maximum amplitude of the turbulence. If the fundamental frequency of the Fermi reflections $\omega(L) = 2\pi c_A / L$ coincides with the frequency of the field (1), then at $h_m = \beta B_0 / 4\sqrt{2}$ the expressions for the diffusion coefficients (6) and (8) do not differ from each other. However, a difference remains in the turbulence spectrum $\Phi(k)$ needed for effective acceleration. In the case of two-dimensional adiabatic variation of the Larmor orbits, a larger turbulence energy is needed for the isotropization of the particle velocities than in the case of the one-dimensional adiabatic Fermi mechanism. Indeed, in the Alfven acceleration ( v= 2) the average amplitude h of pulsations of scale $\lambda$, i.e., the quantity

$$h \sim \left( \int \Phi(k) dk \right)^{1/2} \sim \sqrt{\lambda^{v-1}},$$

decreases with decreasing $\lambda$ more slowly than in the case of Fermi acceleration, when $v > 2$. This leads in the case $v = 2$ to an increase in the role of medium and small scales, i.e., to a stronger scattering, for it is precisely these scales that are responsible for the cyclotron-resonance interaction.

Finally, we present a comparison with the turbulence acceleration investigated in [9]. According to [9], the diffusion coefficient $D_T$ in momentum space in the presence of the spectrum $\Phi(k) \sim k^{-2}$ is given by

$$D_T \approx \frac{c_A^2}{Lc} \left( \frac{h_m}{B_0} \right)^2, \tag{9}$$

so that if

$$\omega(L) = 2\pi c_A / L, \quad h_m = \beta B_0 / 4\sqrt{2},$$

then, according to (7) - (9),

$$D_0 / D_T = D_F / D_T \sim 1/\beta.$$

Thus, at the same form of the spectrum ($v = 2$), the Alfven diffusion coefficient $D_0$ is larger by a factor $1/\beta$ than $D_T$.

## 3. Acceleration of Electrons with Allowance for Synchrotron Radiation: Conditions for the Feasibility of Acceleration

Let us consider an ultrarelativistic stationary problem, in which account is taken of the synchrotron radiation in addition to the Alfven acceleration. When the losses for radiation are taken into account, the diffusion equation (14) of [5] takes the form (see [8])

$$\frac{1}{p^2} \frac{\partial}{\partial p} \left[ p^2 D(p) \frac{\partial \overline{f}}{\partial p} + p^4 \eta \overline{f} \right] = 0, \tag{10}$$

where

$$\eta = \frac{\dot{p}}{p^2} = \frac{4e^4 B_0^2}{9m^3 c^5}, \tag{11}$$

$\dot{p}$ is the rate of momentum loss into synchrotron radiation, averaged over the angle between the field B and p. Equations (10), (15), and (19) of [5] constitute a selfconsistent system of equations for f and $\Phi(k)$, and it is necessary to put $\varepsilon = p$ in (15) of [5], since ultrarelativistic acceleration is being considered.

Choosing as the zeroth approximation $\Phi(k) = \Phi_0 / k^2$ and calculating in accordance with formula (15) of [5] the function D(p) corresponding to this approximation, we find in analogy with the analysis of the preceding section that $D(p) = D_0 p^2$, where $D_0$ is determined from (3). When $D(p) = D_0 p^2$, the solution of (10), leading to a limited plasma density, is

$$\overline{f} = A e^{-\mu p}, \quad \mu = \eta / D_0. \tag{12}$$

It is easy to show that for $\overline{f}$ in the form (12), the second term in the curly bracket of (19) of [5] vanishes, so that in the case of Alfven waves ($\omega = k c_A$) the first approximation for $\Phi(k)$ coincides with the initial zeroth approximation and assumes the universal form (6). Thus, expression (12) is the exact solution of the problem, and, in accordance with (7),

$$D_0 = \frac{\sqrt{2}}{48} \beta W \frac{c_A}{c}.$$

It is possible to present another expression for $D_0$. The intensity of the synchrotron radiation $dJ \sim \overline{f} p^4 dp$, according to (12), has a maximum at $\mu = 4 / p$, so that the quantity $D_0 = \eta / \mu$. can be expressed in terms of the effective radiation frequency $\omega_{eff} \sim \omega_B p^2$ just as in the case of the Fermi acceleration (formula (56) of [8]), namely:

$$D_0 = \frac{\eta p}{4} \approx \frac{1}{9} \frac{m^2 \omega_B^2}{mc^3} \sqrt{\frac{\omega_{eff}}{\omega_B}}. \tag{13}$$

According to expressions (12) and (13) and theresults of [8l, the spectra of the particles in the ultrarelativistic case, with account taken of the Lorentzfriction force, coincides fully for the Alfven and Fermi accelerations.

The constant A in (12) can be expressed in terms of the concentration n of the particles with momentum > p, namely

$$A = n / \int_p^\infty p^2 e^{-\mu p} dp = \frac{n\mu^3}{2\left(1 + \mu p + 1/2 \mu^2 p^2\right) e^{-\mu p}}. \qquad (14)$$

Using (12) and (14), and carrying out the corresponding integration, we obtain the following expressions for the energy density $w_e$ of electrons with momentum > p and for the intensity J of their synchrotron radiation:

$$w_e = Amc^2 \int_p^\infty p^2 e^{-\mu p} dp = \frac{3mc^2 n}{\mu} Q(\mu p), \qquad (15)$$

$$J = Amc^2 \eta \int_p^\infty p^4 e^{-\mu p} dp = \frac{12mc^2 \eta n}{\mu^2} M(\mu p), \qquad (16)$$

where

$$Q(z) = \frac{1 + z + 1/2 z^2 + 1/6 z^3}{1 + z + 1/2 z^2}, \quad M(\mu p) = Q(z) + \frac{1/24 z^4}{1 + z + 1/2 z^2}. \qquad (17)$$

It follows from (15) and (17) that

$$J = \frac{4\eta w_e M(\mu p)}{\mu Q(\mu p)} \sim \eta w_e p, \qquad (18)$$

where account is taken of the fact that zQ(z)/M(z) ~ 4 for all $z = \mu p > 4$.

We note that expressions (15) - (18) are equally valid for the Alfven mechanism and for the Fermi acceleration.

To estimate the minimum concentration nmin of fast particles, at which Alfven acceleration is possible, it is necessary, as noted in Sec. 3, to calculate the increment $\gamma_0(\omega)$ of the resonant oscillations in the absence of scattering. The expression for $\gamma_0(\omega)$ can be obtained by using (16) and (17) of [5] and taking into account the fact that in the absence of scattering the angle part $f_\theta = f_1 + f_2$ of the distribution function (4) of [5] is determined only by the adiabatic variation of the momentum in the alternating field (1). The breakdown of the function $f_\theta$ into $f_1$ and $f_2$ then becomes meaningless, so that the linearized kinetic equation for $f_\theta$, after averaging over the slow period T, can be readily shown to take the form

$$\frac{\partial f_\theta}{\partial t} + \sin^2 \theta \frac{\partial \overline{f}}{\partial p} \frac{\dot{B}}{2B} = 0.$$

Putting $\dot{B}/2B \approx -1/2\beta W \sin Wt$ and substituting the solution of this equation in (17) of [5], we find that

$$\frac{\partial \overline{f}}{\partial p_\perp^{\,2}} - \frac{\partial \overline{f}}{\partial p_z^{\,2}} \frac{\partial \overline{f}}{\partial p}(-\cos Wt + C). \qquad (19)$$

Since the anisotropy (19) should be positive when $\cos Wt \geq 0$, i.e., when $B \geq B_0$, and for fast particles we have $\partial f/\partial p < 0$ on the tail of the distribution, it follows that the integration constant is C = 0. We use for $\overline{f}$ expression (12), which takes into account the synchrotron radiation. Substituting (12) and (19) in (16) of [5], taking (14) into account, and using the dispersion equation of the Alfven waves $\omega = k c_A$, we obtain after integration the following expression for the increment

$$\gamma_0 = A_0 n x e^{-x}(1+x)\left\{\frac{\beta c}{2 c_A x \cos Wt}\left[1 + \frac{x^2 Ei(-x)}{(1+x)e^{-x}}\right] - 1\right\}, \qquad (20)$$

where Ei (-x) is the integral exponential function and where

$$A_0 = \frac{4\pi^3 e^2}{m\omega_B}\left(\frac{c_A}{c}\right)^2\left(1 + \mu p + \frac{\mu^2 p^2}{2}\right)^{-1} e^{-\mu p}, \qquad (21)$$

$$x = \frac{\omega_B}{\omega}\frac{c_A}{c}\mu.. \qquad (22)$$

From the resonance condition

$$p\cos\theta = \omega_B c_A / \omega c$$

we obtain, after the averaging over $\theta$,

$$p^2/3 = (\omega_B c_A / \omega c)^2,$$

so that (40  22) can be written in the form $x = \mu p/3$.

It can be shown that the increment (20) has a maximum at x = 2.3, i.e., at $\mu p = \sqrt{3x} \approx 4$. These results are similar to [11-76]. Within the resonant-frequency band, the momentum changes from $p_0$ to $p_{max}$. By stipulating that $\gamma_0$ be much larger than the frequency W of the alternating field in this band, i.e., by stipulating $n \gg n_{min}$, we can find the value of $n_{min}$ from the equation $\gamma_0(n_{min}, p_{max}) = W$.

We can formulate one more obvious condition for the feasibility of Alfven acceleration, namely:

$$t \geq T, \quad t = \frac{9}{4} \frac{m^3 c^5}{e^4 B_0^2} \sqrt{1 - \frac{v^2}{c^2}} = \frac{1}{\eta p}, \tag{23}$$

where t is the time during which the electron rotating in the field $B_0$ loses an energy equal to its total energy to radiation [10]; p is the dimensionless momentum and $\eta$ is defined by (11). Indeed, in the opposite case the particle loses its entire energy to radiation before isotropization of the momenta takes place, so that the Alfven cycle does not lead to acceleration.

## References

1. H. Alfven, Phys. Rev., 75, 1732 {1949).

2. H. Alfven, E. Astrom., Nature, 181, 330 {1958).

3. H. Alfven and K.-H. Felthammer, Cosmic Electrodynamics, {Russian translation), Mir, (1967), M. F. Bakhareva, V. N. Lomonosov, B. A. Tverskoi, Zh. Eksp. Teor. Fiz. 59, 2003 (1970).

4. A. Schluter, Zs. Naturforsch, 12a, 822 {1957).

5. M.A. Kutlan, arXiv:1801.00447 (2018).

6. A. A. Vedenov, Voprosy teorii plazmy (Problems of Plasma Theory), No.3, 203 (1963).

7. R. Z. Sagdeev and V. D. Shafranov, Zh. Eksp. Teor. Fiz. 39, 181 {1960).

8. B. A. Tversko'l, Zh. Eksp. Teor. Fiz., 52, 483 {1967).

9. B. A. Tversko'i, Zh. Eksp. Teor. Fiz., 53, 1417 {1967).

10. J. G. Scargle, Astron. Journ., 156, 401 {1969).

11. Fedorov M.V., Oganesyan K.B., Prokhorov A.M., Appl. Phys. Lett., **53**, 353 (1988).

12. Oganesyan K.B., Prokhorov A.M., Fedorov M.V., Sov. Phys. JETP, **68,** 1342 (1988).

13. Oganesyan KB, Prokhorov AM, Fedorov MV, Zh. Eksp. Teor. Fiz., **53**, 80 (1988).

14. E.A. Ayryan, K.G. Petrosyan, A.H. Gevorgyan, N.Sh. Izmailian, K.B. Oganesyan, arXiv:1703,00813 (2017).

15. Edik A. Ayryan, Karen G. Petrosyan, Ashot H. Gevorgyan, Nikolay Sh. Izmailian, K. B. Oganesyan, arXiv:1702.03209 (2017).

16. Ashot H. Gevorgyan, K. B. Oganesyan, Edik A. Ayryan, Michal Hnatic, Yuri V. Rostovtsev, Gershon Kurizki, arXiv:1703.07637 (2017).

17. L.A.Gabrielyan, Y.A.Garibyan, Y.R.Nazaryan, K.B. Oganesyan, M.A.Oganesyan,

M.L.Petrosyan, A.H. Gevorgyan, E.A. Ayryan, Yu.V. Rostovtsev, arXiv:1701.00916 (2017).

18. M.L. Petrosyan, L.A. Gabrielyan, Yu.R. Nazaryan, G.Kh.Tovmasyan, K.B. Oganesyan, A.H. Gevorgyan, E.A. Ayryan, Yu.V. Rostovtsev, arXiv:1704.04730.

19. A.S. Gevorkyan , K.B. Oganesyan , E.A. Ayryan , Yu.V. Rostovtsev, arXiv:1706.03627 (2017).

20. A.S. Gevorkyan, K.B. Oganesyan, E.A. Ayryan, Yu.V. Rostovtsev, arXiv:1705.09973 (2017).

21. D.N. Klochkov, A.H. Gevorgyan, K.B. Oganesyan, N.S. Ananikian, N.Sh. Izmailian, Yu. V. Rostovtsev, G. Kurizki, arXiv:1704.06790 (2017).

22. K.B. Oganesyan, J. Contemp. Phys. (Armenian Academy of Sciences), **52**, 91 (2017).

23. AS Gevorkyan, AA Gevorkyan, KB Oganesyan, Physics of Atomic Nuclei, **73**, 320 (2010).

24. D.N. Klochkov, AI Artemiev, KB Oganesyan, YV Rostovtsev, MO Scully, CK Hu, Physica Scripta, **T140**, 014049 (2010).

25. K.B. Oganesyan, M.L. Petrosyan, YerPHI-475(18) – 81, Yerevan, (1981).

26. AH Gevorgyan, MZ Harutyunyan, KB Oganesyan, MS Rafayelyan, Optik-International Journal for Light and Electron, Optics, 123, 2076 (2012).

27. D.N. Klochkov, AI Artemiev, KB Oganesyan, YV Rostovtsev, CK Hu, J. of Modern Optics, **57,** 2060 (2010).

28. K.B. Oganesyan, J. Mod. Optics, **62,** 933 (2015).

29. K.B. Oganesyan. Laser Physics Letters, **12**, 116002 (2015).

30. GA Amatuni, AS Gevorkyan, AA Hakobyan, KB Oganesyan, et al, Laser Physics, **18,** 608 (2008).

31. K.B. Oganesyan, J. Mod. Optics, **62,** 933 (2015).

32. Petrosyan M.L., Gabrielyan L.A., Nazaryan Yu.R., Tovmasyan G.Kh., Oganesyan K.B., Laser Physics, **17**, 1077 (2007).

33. AH Gevorgyan, KB Oganesyan, EM Harutyunyan, SO Arutyunyan, Optics Communications, **283**, 3707 (2010).

34. E.A. Nersesov, K.B. Oganesyan, M.V. Fedorov, Zhurnal Tekhnicheskoi Fiziki, **56**, 2402 (1986).

35. A.H. Gevorgyan, K.B. Oganesyan, Optics and Spectroscopy, **110**, 952 (2011).

36. A.H. Gevorgyan, M.Z. Harutyunyan, K.B. Oganesyan, E.A. Ayryan, M.S. Rafayelyan, Michal Hnatic, Yuri V. Rostovtsev, G. Kurizki, arXiv:1704.03259 (2017).

37. K.B. Oganesyan, J. Mod. Optics, **61,** 763 (2014).

38. A.H. Gevorgyan, K.B.Oganesyan,  E.M.Harutyunyan, S.O.Harutyunyan,  Modern Phys. Lett. B, **25**, 1511 (2011).

39. A.H. Gevorgyan**,** M.Z. Harutyunyan, G.K. Matinyan, K.B. Oganesyan, Yu.V. Rostovtsev, G. Kurizki and M.O. Scully**,**  Laser Physics Lett., **13,** 046002 (2016).

40. A.I. Artemyev, M.V. Fedorov, A.S. Gevorkyan, N.Sh. Izmailyan, R.V. Karapetyan, A.A. Akopyan, K.B. Oganesyan, Yu.V. Rostovtsev, M.O. Scully, G. Kuritzki, J. Mod. Optics, **56**, 2148 (2009).

41. A.S. Gevorkyan, K.B. Oganesyan, Y.V. Rostovtsev, G. Kurizki, Laser Physics Lett., **12**, 076002 (2015).

42. K.B. Oganesyan, J. Contemp. Phys. (Armenian Academy of Sciences),  **50,** 312 (2015).

43. ZS Gevorkian, KB Oganesyan, Laser Physics Letters **13**, 116002 (2016).

44. AI Artem'ev, DN Klochkov, K Oganesyan, YV Rostovtsev, MV Fedorov, Laser Physics **17**, 1213 (2007).

45. Zaretsky, D.F., Nersesov, E.A., Oganesyan, K.B.,  Fedorov, M.V., Sov. J. Quantum Electronics, **16**, 448 (1986).

46. K.B. Oganesyan,  J. Contemp. Phys. (Armenian Academy of Sciences), **50,** 123 (2015).

47. DN Klochkov, AH Gevorgyan, NSh Izmailian, KB Oganesyan, J. Contemp. Phys., **51,** 237 (2016).

48. K.B. Oganesyan, M.L. Petrosyan, M.V. Fedorov, A.I. Artemiev, Y.V. Rostovtsev, M.O. Scully, G. Kurizki, C.-K. Hu,  Physica Scripta, **T140**, 014058 (2010).

49. AS Gevorkyan, AA Gevorkyan, KB Oganesyan, GO Sargsyan, Physica Scripta, **T140,** 014045 (2010).

50. AH Gevorgyan, KB Oganesyan, Journal of Contemporary Physics (Armenian Academy of Sciences) **45,** 209 (2010).

51. K.B. Oganesyan, J. Mod. Optics, **61,** 1398  (2014).

52. AH Gevorgyan, KB Oganesyan, GA Vardanyan, GK Matinyan, Laser Physics, **24,** 115801 (2014)

53. K.B. Oganesyan,  J. Contemp. Phys. (Armenian Academy of Sciences),  **51,** 307 (2016).

54. AH Gevorgyan, KB Oganesyan, Laser Physics Letters **12** (12), 125805 (2015).

55. Oganesyan K.B., Prokhorov, A.M.,  Fedorov, M.V., ZhETF, **94**, 80 (1988).

56. E.M. Sarkisyan, KG Petrosyan, KB Oganesyan, AA Hakobyan, VA Saakyan, Laser Physics, **18,** 621 (2008).

57. A.H. Gevorgyan, K.B. Oganesyan, R.V. Karapetyan, M.S. Rafaelyan,  Laser Physics Letters,  **10**, 125802 (2013).

58. K.B. Oganesyan, Journal of Contemporary Physics (Armenian Academy of Sciences) **51,** 10 (2016).

59. M.V. Fedorov, E.A. Nersesov, K.B. Oganesyan, Sov. Phys. JTP, **31,** 1437 (1986).

60. K.B. Oganesyan, M.V. Fedorov, *Zhurnal Tekhnicheskoi Fiziki*, **57**, 2105 (1987).

61. M.V. Fedorov, K.B. Oganesyan, IEEE J. Quant. Electr, **QE-21**, 1059 (1985).

62. D.F. Zaretsky, E.A. Nersesov, K.B. Oganesyan, M.V. Fedorov, Kvantovaya Elektron. **13** 685 (1986).

63. A.H. Gevorgyan, K.B. Oganesyan, E.A. Ayryan, M. Hnatic, J.Busa, E. Aliyev, A.M. Khvedelidze, Yu.V. Rostovtsev, G. Kurizki, arXiv:1703.03715 (2017).

64. A.H. Gevorgyan, K.B. Oganesyan, E.A. Ayryan, Michal Hnatic, Yuri V. Rostovtsev, arXiv:1704.01499 (2017).

65. E.A. Ayryan, A.H. Gevorgyan, K.B.Oganesyan, arXiv:1611.04094 (2016).

66. A.S. Gevorkyan, K.B. Oganesyan, E.A. Ayryan, Yu.V. Rostovtsev, arXiv:1705.09973 (2017).

67. A.H. Gevorgyan, K.B. Oganesyan, Laser Physics Letters, **15**, 016004 (2018).

68. K.B. Oganesyan, arXiv:1611.08774 (2016).

69. E.A. Ayryan, A.H. Gevorgyan, N.Sh. Izmailian, K.B. Oganesyan, arXiv:1611.06515 (2016).

70. I.V. Dovgan, K.B. Oganesyan, arXiv:1612.04608v2 (2016).

71. V.V. Arutyunyan, N. Sh. Izmailyan, K.B. Oganesyan, K.G. Petrosyan and Cin-Kun Hu, Laser Physics, **17**, 1073 (2007).

72. E.A. Ayryan, M. Hnatic, K.G. Petrosyan, A.H. Gevorgyan, N.Sh. Izmailian, K.B. Oganesyan, arXiv: 1701.07637 (2017).

73. DN Klochkov, KB Oganesyan, EA Ayryan, NS Izmailian, Journal of Modern Optics **63,** 653 (2016).

74. K.B. Oganesyan. Laser Physics Letters, **13**, 056001 (2016).

75. DN Klochkov, KB Oganesyan, YV Rostovtsev, G Kurizki, Laser Physics Letters **11,** 125001 (2014).

76. K.B. Oganesyan, Nucl. Instrum. Methods A **812,** 33 (2016).